\documentclass[aps,pre,twocolumn,showpacs,superscriptaddress,groupedaddress]{revtex4}
\usepackage{graphicx,subfigure}
\usepackage{hyperref}
\usepackage[svgnames]{xcolor}
\usepackage{amsmath}
\usepackage{amsfonts}
\usepackage{bm}
\usepackage{amssymb}
\usepackage{varioref}
\usepackage{float}
\usepackage{dcolumn}   
\usepackage{subfigure}
\usepackage{array}
\usepackage{tikz}
\usepackage{pgfplots}
\usepackage{mathtools}
\graphicspath{{./Figures/}}
\usetikzlibrary{positioning,fadings,through}
\definecolor{left} {HTML}{001528}

\begin{document}
\title{A phase separation of  active colloidal suspension via  Quorum-sensing}
\author{Francis Jose}
\email{francisjm24@gmail.com}
\author{Shalabh K. Anand }
\email{skanand@iiserb.ac.in}
\author{Sunil P. Singh}
\email{spsingh@iiserb.ac.in}
%\noaffiliation
\affiliation{Department of Physics,\\ Indian Institute of Science Education and Research, \\Bhopal 462 066, Madhya Pradesh, India}

\begin{abstract}
	We present the Brownian dynamics simulation of active colloidal suspension in two dimensions, where the self-propulsion speed of a colloid is regulated according to the local density sensed by it. The role of concentration-dependent motility on the phase-separation of colloids and their dynamics is investigated in detail. Interestingly, the system phase separates at a very low packing fraction ($\Phi\approx 0.125$) at higher self-propulsion speeds ($\text{Pe}$),  which coexists with a homogeneous phase and attains long-range crystalline order beyond a transition point. The transition point is quantified here from the local density profiles,   local and global-bond order parameters. We have shown that the phase diagram's characteristics are qualitatively akin to the active Brownian particle (ABP) model.  Moreover,  our investigation reveals that the density-dependent motility amplifies the slow-down of the directed speed,  which facilitates phase-separation even at low packing fractions. The effective diffusivity shows a crossover from quadratic rise to a power-law behavior of exponent $3/2$  with $\text{Pe}$ in the phase-separated regime. Furthermore, we have shown that the effective diffusion decreases  exponentially  with packing fraction in the phase-separated regime while  linear decrease in the single phase regime.
\end{abstract}	

\maketitle

\section{ Introduction} 
A collection of self-propelled units constitutes an out-of-equilibrium system that harvests ambient energy from the environment to perform a long-persistent motion~\cite{marchetti2013hydrodynamics,cates2012diffusive,zottl2016emergent}. These systems exhibit  complex collective dynamics which are markedly different from their passive counterpart, such as self-organisation at low density ~\cite{corte2008random,peruani2006nonequilibrium,gopinath2012dynamical,pohl2014dynamic,lin2018collective,Narayan105,PhysRevLett.96.180602,reentrant_baskaran}, flocking~\cite{mora2016local,ballerini2008interaction}, swarming~\cite{cavagna2017dynamic}, vortex-formation~\cite{bricard2015emergent}, the formation of linear chains~\cite{yan2016reconfiguring},
anomalously large density fluctuations~\cite{fily2012athermal}, non-monotonic behavior in the active pressure~\cite{winkler2015virial,swim_kardar,Brady_swim_prl2014}, etc. In the past,  phase separation of the self-propelled systems has been rigorously explored in the laboratories where  dynamical structures are contrived by employing light controlled motion of active colloids~\cite{palacci2013living,bricard2015emergent,singh2017non}, diffusiophoretic motion or concentration-gradient driven motion~\cite{buttinoni2013dynamical,buttinoni2012active,ginot2018aggregation,singh2017non,theurkauff2012dynamic,palacci2014light,schmidt2019light}.

Within the framework of theory and simulations, the phase-separation of a self-propelled system of  hard colloids  has been explored  by treating their motion  either as a `run and tumble' model~\cite{lee2019computational,berg1972chemotaxis,turner2000real,ariel2018collective}, active Ornstein-Uhlenbeck (AOU) particles~\cite{paoluzzi2016critical,maggi2020universality,caprini2019comparative}, or as an active Brownian particle (ABP) model~\cite{bialke2012crystallization,speck2014effective,speck2015dynamical,Bialk__2013,C7SM00519A,Cates11715,baskar_quorum,redner2013structure,anand2018structure,anand2020conformation}. The self-propulsion causes phase separation referred as motility-induced phase separation (MIPS)~\cite{bialke2015active,cates2015motility,cates2013active,digregorio2018full,gonnella2015motility,barre2015motility,suma2014motility,sese2018velocity,fily2012athermal,redner2013structure,Cates_2013,PhysRevLett.121.098003,caprini2020spontaneous,solon2018generalized}, which is a consequence of the slow-down of the  active colloids produced by steric-interactions~\cite{patch2017kinetics,solon2018generalized}. In MIPS, a liquid-like dense aggregate of slow-moving colloids coexists with  a  gas-like phase of fast-moving colloids at low-density~\cite{bialke2015active,cates2015motility,cates2013active,digregorio2018full,gonnella2015motility,barre2015motility,suma2014motility,sese2018velocity,fily2012athermal,redner2013structure,Cates_2013,PhysRevLett.121.098003,solon2018generalized}. In the standard ABP model, the phase-separation arises at packing fractions  ($\Phi=0.4$) even in the absence of any attractive or alignment interactions~\cite{fily2012athermal}.  Additionally, the presence of an attractive short-range interaction among ABPs causes significant changes on local inhomogeneity, separation of mixtures, and also adds to re-entrant phase behavior~\cite{reentrant_baskaran,agudo2019active}.

%The collective dynamics of self-propelled repulsive hard-colloids is extensively studied as it exhibits rich dynamical phase behaviour ~\cite{marchetti2013hydrodynamics,cates2012diffusive,redner2013structure,fily2012athermal}.  However, density-dependent motility enhances the richness of the phase diagram of repulsive ABP systems distinctively, which will be prime focus of this study.}

%The investigations in the area of active colloids can be mainly categorised into two themes, one is  alignment interaction induced ordering~\cite{vicsek1995novel,toner1995long,vicsek2012collective,chate2008modeling,toner2012reanalysis,fily2012athermal,redner2013structure}, 

 %The alignment interaction induces polar ordering as a consequence of that system displays ordering in the structure and velocity fields   ~\cite{vicsek1995novel,mahault2018self,gregoire2004onset,solon2015phase,caussin2014emergent,peruani2006nonequilibrium,aranson2003model,ginelli2010large,deseigne2010collective}. 

 The aggregation in biological systems at low density is a ubiquitous phenomenon~\cite{ishimoto2018hydrodynamic,reid1990bacterial,schramm2020protein}. Concurrently, coordinated motion in these systems is often a consequence of complex communication pathways among their individual units. Bacterial colonies, like {\it Aliivibrio fischeri, Vibrio harveyi, Erwinia carotovora}, etc., produce extracellular enzymes called auto-inducers, which regulate their gene expression~\cite{fuqua1996census,brown2001cooperation,lupp2005vibrio,sperandio2002quorum,bassler1999bacteria,crespi2001evolution},  thus act as a means of communication among cells~\cite{bauerle2018self,bassler1999bacteria}. For instance, \textit{Aliivibrio fischeri}~\cite{lupp2005vibrio} produces luminescence after  their local population growth surpasses a threshold density. The communication among the bacterial cells triggered  by the variation in local population density is termed   `quorum-sensing'.  It plays vital role in the regulation of various physiological processes like  manoeuvring local population density, regulation of gene expression~\cite{lupp2005vibrio,sperandio2002quorum,bassler1999bacteria}, communication among cells~\cite{bauerle2018self,bassler1999bacteria},  motility~\cite{bauerle2018self,sperandio2002quorum}, etc. Despite the vast applications of  `quorum-sensing' in biology, this subject has been  addressed far less  in simulations,  especially  for the ABPs~\cite{baskar_quorum,velasco2018collective,solon2018generalized}.
 
  The variation of self-propulsion speed as a function of local density has been accounted  in theoretical continuum models using  phenomenological hydrodynamic equations with local density as a coarse-grained field. Interestingly, this model  also unveils  MIPS. Importantly, it provides a unified approach for the phase-separation using quorum-sensing rule and the generic constant activity model with  pair-wise interactions ~\cite{farrell2012pattern,solon2018generalized,cates2015motility}.
 A detailed study on the role of controlled and explicit density-dependent activity on the MIPS would provide more insights into the underlining behavior. We attempt here to reveal  the role of quorum-sensing contribution along with pair-wise repulsive forces in the phase separation. Additionally, we compare our results to that of generic ABP model.  %The interaction range of quorum-sensing is assumed larger than the pair-wise interaction potential. {\cred Last sentence seems to be irrelevant here.} } 
  
%{\cred  On the contrary,  density-dependent propulsion is often adopted in research involving continuum models~\cite{farrell2012pattern,solon2018generalized,cates2015motility}. We attempt to broaden the present understanding of the impact density dependant motility has on the  phase separation and compare it to generic MIPS in ABPs.}  

%In this article, we consider the role of density-dependent motility and compare its insight on the  phase-separation of colloids  which has not been taken into account in previous simulations.

%A `quorum-sensing' model, probes the propulsion scheme to regulate the motility with respect to their local density,  displays a rich phase diagram for rod-like colloids~\cite{baskar_quorum,velasco2018collective}. { In general, such models lack effective momentum conservation~\cite{solon2018generalized,fily2017mechanical}, and has been studied using theoretical tools~\cite{farrell2012pattern,solon2018generalized}. 

The article presents  Brownian dynamics simulation of ABPs whose motility is regulated by linearly decreasing  self-propulsion force with local density. The density-dependent motility induces cluster formation with hexatic ordering at relatively  low  packing fractions $\Phi \geq 0.125$, in the high activity regime. The local density distribution confirms the presence of large dense clustered phase coexisting with a homogeneous low-density phase beyond a critical density. The local and global order parameters such as $\Psi_6$ and $C_{q_6}$, respectively, ascertain the presence of hexatic ordering at higher P{\'e}clet numbers.  We  have  shown  that  the  emergence  of  the coexistence of low and high-density phases at very low packing is a consequence of multi-body collisions leading to drastic slow-down of the swim speed.  %{\cred The emergence of phase separation at very low densities is a result of the drastic slow down of the directed speed of ABPs at high activity strength.} 
%{ In addition, the mean-squared-displacement (MSD) of colloids yields a rich dynamical behavior with a sub-diffusion in the short-time limit followed by a plateau and a super-diffusive regime in the intermediate time window. }
{ In addition, the effective diffusion coefficient  in the phase-separated state displays a cross over from a parabolic rise to a power-law behavior with an exponent $3/2$ with P{\'e}clet number~\cite{winkler2015virial,Brady_swim_prl2014}. Moreover, the total pressure exhibits a sharp decrease in the phase-separated regime, as expected.}

The article is organised as follows: Section 2 elaborates the simulation approach, all the results are presented in section 3 that consists 5 different sub-sections,  with the summary in section 4.    

\section{Simulation Model} 
 We model self-propelled discs,  in two dimensions, interacting with each other via repulsive-shifted Lennard Jones (LJ) potential,
\begin{equation}
    U_{LJ}(r_{ij})=
    \left \{
    \begin{aligned}
    &4\epsilon\Big[\big(\frac{\sigma}{r_{ij}}\big)^{12}-\big(\frac{\sigma}{r_{ij}}\big)^{6}\Big]+\epsilon, && \ r_{ij}\leq 2^{1/6}\sigma,\\ 
    &0, && \  r_{ij}>2^{1/6}\sigma.
    \end{aligned} \right.
    \label{Eq:lj}
\end{equation}
Where $r_{ij}=|{\bf r}_{i}-{\bf r}_{j}|$ is the distance between the pair $i$ and $j$, $\epsilon$ is the LJ energy, and $\sigma$ is the diameter of  disc. 

The equation of motion for the position and its orientation vector are governed by the over-damped Langevin equation,
\begin{eqnarray}\nonumber
    \bf{\Dot{r}}_i&=&\frac{1}{\gamma}\left[-\sum_{j=1}^{N_m}{\bf \nabla}_{i}U_{LJ}({r}_{ij})+F_{a,i}^{'}{\bf{\hat e}}_{i}+\bf{F}_{i}^{R}\right],\\
    \Dot{\theta_{i}}& = & F_{i}^{\theta}. 
    \label{motion}	
\end{eqnarray}
Where $\gamma$ is the drag coefficient, $F_{i}^{R}$ is the thermal noise with zero mean, $F_{a,i}^{'}$ is the magnitude of the active force, and its orientation is given by the unit vector $\hat {\bf e}_i=(\cos\theta_i,\sin \theta_i)$. The direction of polar  vector $\hat {\bf e}$ obeys the equation of motion of rotational diffusion, angle $\theta_{i}$ measured from the x-axis.

The friction coefficient and the thermal noise are coupled via fluctuation-dissipation relation, $\big \langle F_{i\alpha }^{R}(t)F_{j{\beta}}^{R}(t{'})\big \rangle=2\gamma\delta_{\alpha,\beta}\delta_{ij} k_{B}T\delta(t-t^{'})$. Similarly, $F_{i}^{\theta}$ is a random torque on a particle, and the relation of random torque with rotational diffusion is given as $\big \langle F_{i}^{\theta}(t).F_{j}^{\theta}(t^{'})\big \rangle=2D_{R}\delta(t-t^{'})\delta_{ij}$. Where, $D_{t} = k_{B}T/\gamma$ and $D_{R}=3D_{t}/\sigma^{2}$ are the translational and rotational diffusion constants, respectively.

The effect of local environment on the motility is taken  in an approximate manner. For simplicity, motility ($F_{a,i}^{'}$) of a particle is assumed to be a linear function of its nearest neighbors within a fixed cut-off distance, $R_c \leq 1.3$. { Thus the active force linearly diminishes with the increase in local population as,
\begin{equation}
    F_{a,i}^{'}=
    \left \{
    \begin{aligned}
    &F_{a,i}, &&\; n \le 2,\\ 
    &\beta F_{a,i}\Big(\alpha-\frac{\phi_{i}}{\phi_{m}}\Big), &&\; 2 < n\leq 6,
    \end{aligned}
    \right.
    \label{Eq:qur}
\end{equation}
 and $ F_{a,i}^{'}=0$ for $n > 6$, where $n$ stands for the number of nearest neighbors of the $i^{th}$ particle, $F_{a,i}$ is the active force on an isolated particle that is modified to $F_{a,i}^{'}$ in the presence of  neighbors within a cut-off distance of $R_{c}=1.3\sigma$.} Here,  $R_c$ is chosen  approximately  to the first minimum in the radial distribution function of a dense system. The cut-off is chosen larger than the pair-wise interaction potential. %A larger cut-off  also reveals qualitatively quite alike physical behavior. } 

The constants $\alpha=1$ and $\beta\approx1.8$ are chosen so that the function is calibrated in a linearly decreasing form from $F^{'}_{a,i}=F_{a,i}$ to $0$. If the center of a neighboring particle crosses the cut-off distance, then it is assumed to be completely inside the screening area. The apparent local area fraction of the circle of radius $R_c$ within the $n$ neighbors of particle $i$ and itself at the center is approximated as $\phi_{i}=\lambda\big[\pi(\frac{\sigma}{2})^2+n\mathcal{A}\big]$, with $\lambda=(\pi R_c^2)^{-1}$, $\mathcal{A}\approx0.6531\sigma^2$ is the approximate maximum area of the neighboring disk to $i$ located within the cut-off distance $R_c$, $\phi_{m}=\lambda\big[\pi(\frac{\sigma}{2})^2+n\mathcal{A}\big]$ which is same as $\phi_i$ at $n=6$.

The parameters  are presented in the dimensionless form with the length scaled by $\sigma$, energy in the unit of $k_BT$ (thermal energy), and time in  unit of $\tau=\sigma^{2}/D_t$. The Brownian dynamics simulation is employed with a time step in the range of $10^{-4}\tau$ to $10^{-5}\tau$. The simulation box is taken to be a square with a side of length $L$ and periodic in each direction. All the results are quantified in the parameter landscape of packing fraction $\Phi=\pi\sigma^{2}N/4L^2$, and dimensionless P{\'e}clet number $\text{Pe} = F_{a}\sigma/k_{B}T$ in equivalence with self-propulsion speed. The number of particles  $N=5041$, unless stated otherwise. The box length $L$ is varied to achieve the packing fractions $0.05$ to $0.4$ and $\text{Pe}$ is varied in the range of $0$ to $250$. The good statistics for each simulation data set is generated by averaging over ten independent ensembles.

\begin{figure}
 	\includegraphics[width=\linewidth]{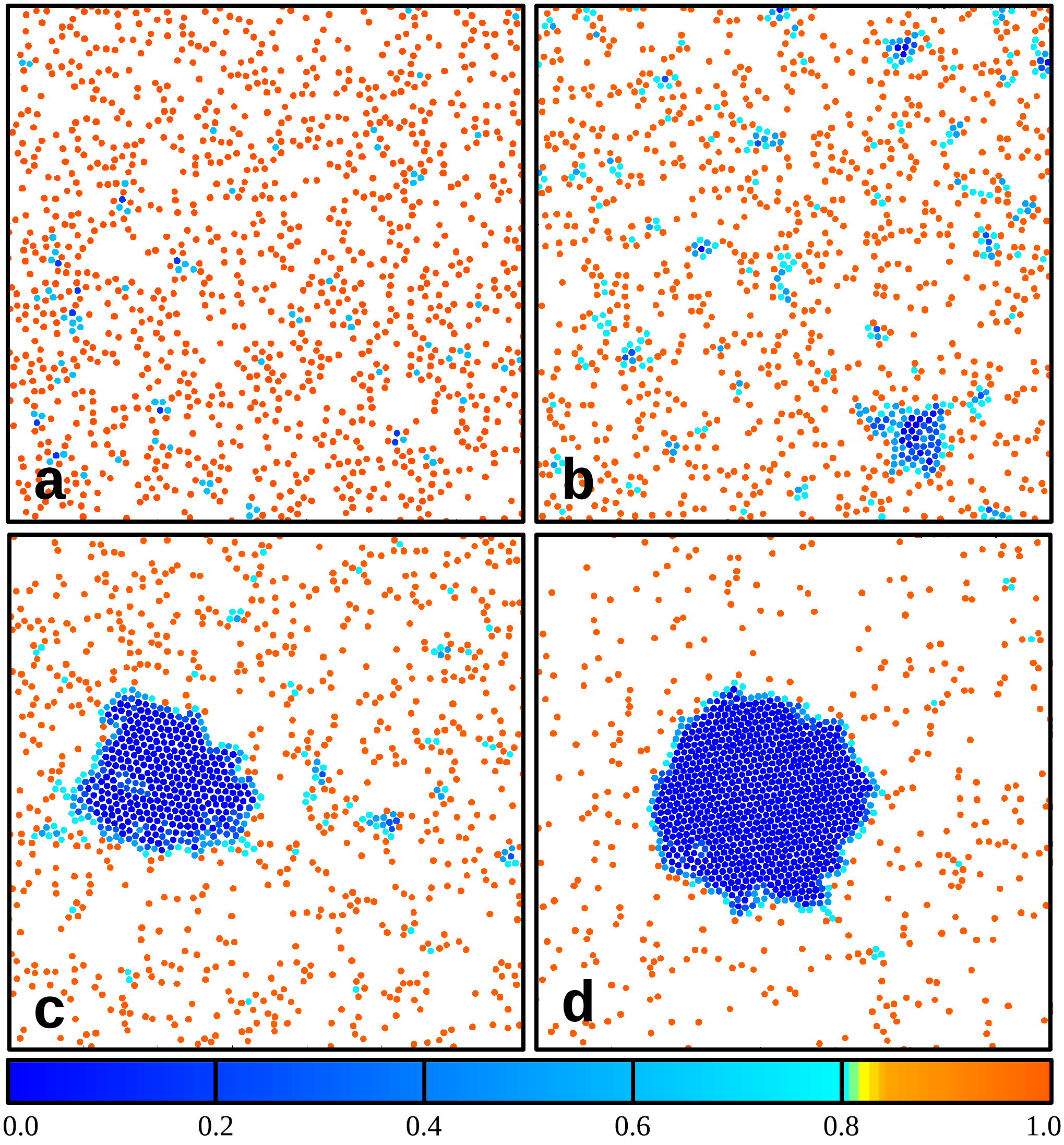}
    \caption{A set of snapshots corresponding to various P{\'e}clet numbers, a) $\text{Pe}=5$, b) $\text{Pe}=50$, c) $\text{Pe}=60$, and d) $\text{Pe}=200$  at $\Phi=0.20$, $N=1225$ . The particles are color-coded according to their fraction of self-propulsion force $F_{a,i}^{'}/F_{a,i}$. The color gradient denotes the relative speed of the particles from fast motion (red) to slow motion (blue).}
	\label{Fig:snapshot}
\end{figure}
  
\begin{figure}[t]
    \includegraphics[width=\linewidth]{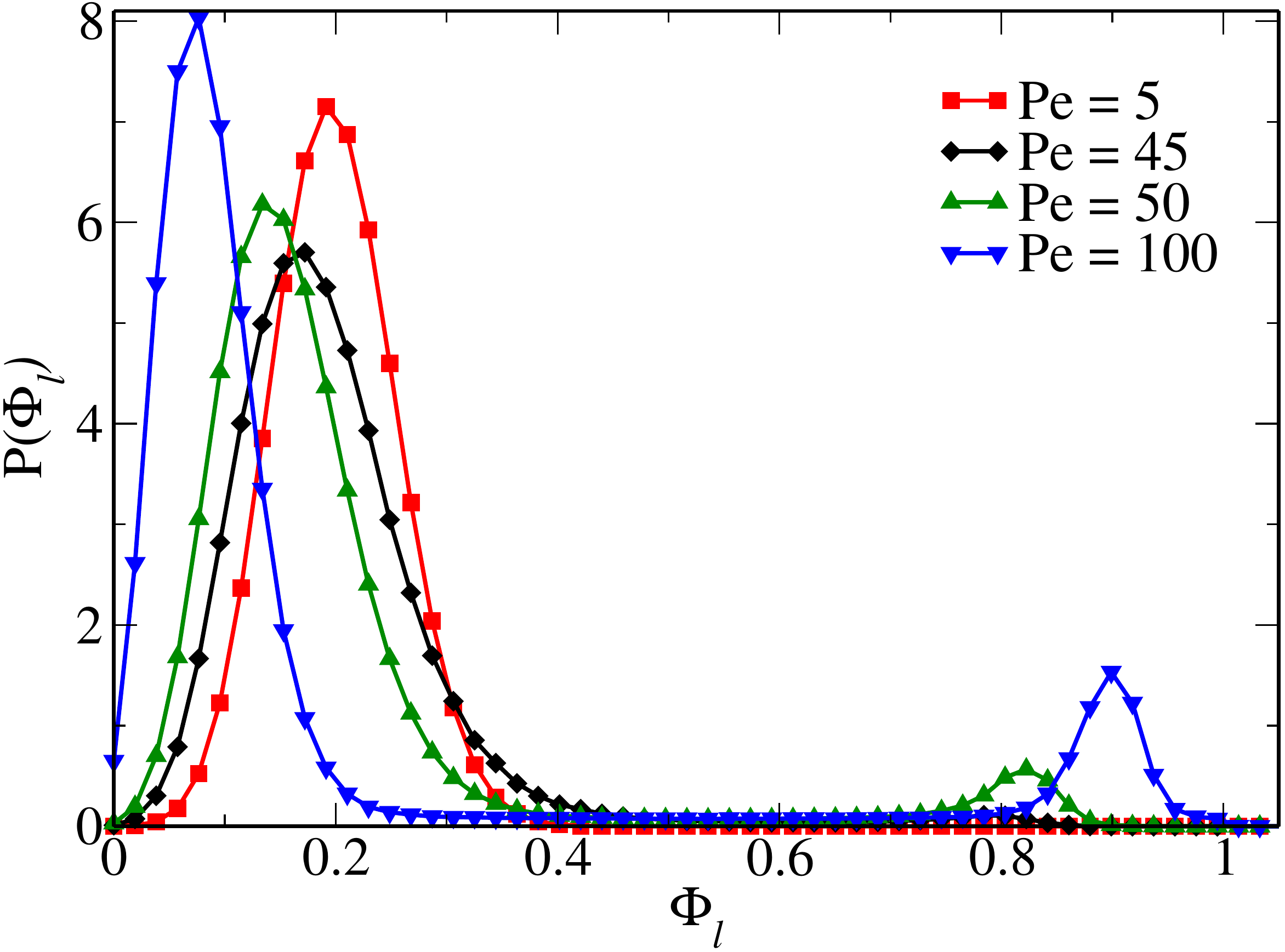}
	\includegraphics[width=\linewidth]{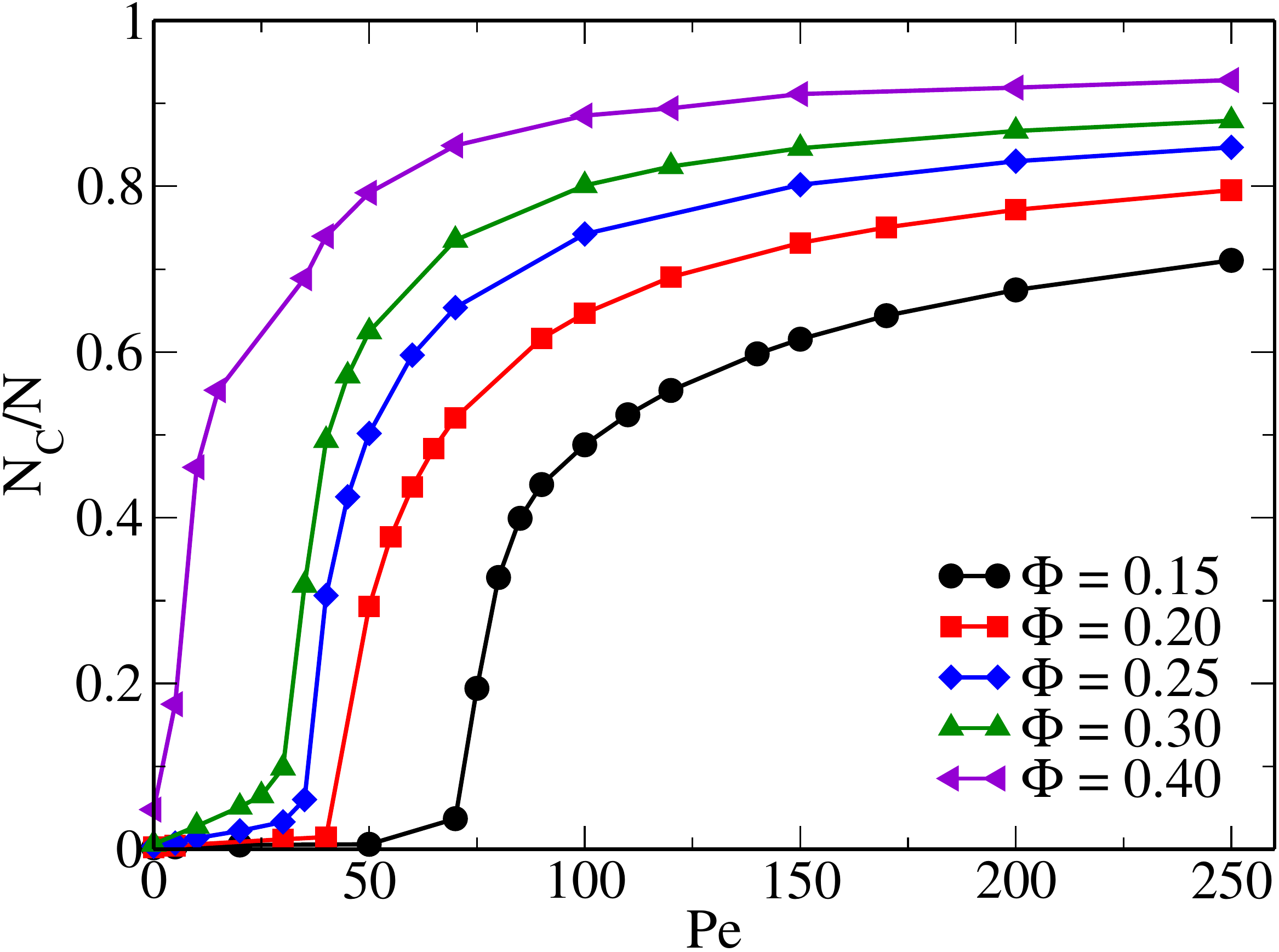}
		\caption{% a) The cluster size distribution $P(N_S)$ for various $\text{Pe}$ in the range $0$ to $250$ for $\Phi=0.2$. The inset shows the distribution of smaller clusters of size ($N_S$) in the range of $0$ to $20$ at $\Phi=0.2$. Note that this is not shown in the main plot.
	 a) The probability distribution $P(\Phi_l)$ of the local packing fractions plotted for various Pe at given $\Phi=0.2$. The curve becomes bimodal at the transition point $\text{Pe}\approx 45$ depicting the presence of a dense clustered region.
	b) The average fractional size of the largest cluster $N_C/N$ ( $N_C$ is average size of the largest cluster),  with $\text{Pe}$ for various packing fractions ($\Phi$) in the range of $0.15$ to $0.4$. }
	\label{Fig:clustersize}
\end{figure}

\section{Results}
%{ The collective dynamics of self-propelled repulsive hard-colloids is extensively studied as it exhibits rich dynamical phase behaviour ~\cite{marchetti2013hydrodynamics,cates2012diffusive,redner2013structure,fily2012athermal}. The presence of an attractive short-range interaction among ABPs causes significant changes on local inhomogeneity, separation of mixtures, and also leads to re-entrant phase behavior~\cite{reentrant_baskaran,agudo2019active}. However, density-dependent motility enhances the richness of the phase diagram of repulsive ABP systems distinctively, which will be prime focus of this study.}% for the case of repulsive colloids.}

\subsection{Phase Separation}
  The considered system of quorum sensing ABPs is observed to display phase-separation at a high activity strength even at very low densities. The simulation snapshots in Fig.~\ref{Fig:snapshot}- a, b, c, and d, at  $\text{Pe}=5$, $50$, $60$, and $200$ respectively, illustrate the emergence of the aggregated-dense structures. For low P{\'e}clet numbers, the system exhibits mostly gas-like homogeneity, with a few small aggregates having a short life-time.  { These aggregates grow in the intermediate  regime ($5<\text{Pe}<50$ for $\Phi=0.2$) with liquid-like characteristics (see Fig.~\ref{Fig:snapshot}-a and b).} However, the dense aggregate grows further for large $\text{Pe}$ and remains stable for all time scales. The size of the largest aggregate attains the order of the system size in the asymptotic limit  $\text{Pe}>>50$ for $\Phi = 0.2$ (see  Fig.~\ref{Fig:snapshot}-d). Moreover, once in the steady-state at high P{\'e}clet numbers, the aggregate acquires a long-range ordering, as Fig.~\ref{Fig:snapshot}-c and d illustrate, concurrent with the previous studies~\cite{redner2013structure,fily2012athermal}.   

 The phase separation in ABPs is a consequence of multi-body collisions at large $\text{Pe}$ causing slow-down of the speed in the denser region, which further leads to the aggregation of slow-moving colloids. Figure~\ref{Fig:snapshot} illustrates this effect, where blue color represents slow-moving colloids while the reds are fast-moving.
{  The chosen model for density-dependent motility imparts  influence  on microscopic dynamics. Hence, it is imperative to examine the structural properties  from a microscopic perspective. For that, we  probe the heterogeneous local density profiles appearing at large $\text{Pe}$. In order to compute the normalised local density profile (Fig.\ref{Fig:clustersize}-a), the simulation box is divided into smaller subsystems  ($6 \times 6$), which is approximately 500 times smaller than the original system. The local packing fraction $\Phi_l$ is estimated in these subsystems at $\Phi=0.2$. In the large $\text{Pe}$ limit ($\text{Pe}\geq50$ for $\Phi = 0.2$), the normalised density distribution displays a bimodal peak, one at a lower density and the other at higher density (aggregated phase). The characteristic feature of phase-separation, i.e. coexistence of low and high density phases, is distinctly captured in the local density profile beyond a transition  P{\'e}clet number termed here onward as $\text{Pe}_c$.}

{ The distribution $P(\Phi_l)$ suggests that beyond the transition point, the system also consists of a dense aggregate, which we call the largest cluster ($N_C$).}   The growth of the clustered phase is quantified using $N_C=\left<\max\{N_S\}\right>$, where $N_S$ is the size of any cluster in the simulation box.  Fig.~\ref{Fig:clustersize}-b illustrates  $N_C/N$ as a function of $\text{Pe}$ for a range of $\Phi=0.15$ to $0.4$.   {  The presence of a prominent dense clustered phase results in a sharp transition in the largest cluster size and its size tends to saturate in the limit of high Pe far beyond the transition point.}

\subsection{Structure of  Cluster}
%Long-range ordering with a hexagonal crystalline structure is prevalent in the largest aggregate for higher activity. To quantify the internal structure and its ordering in 2D, we analyse an average angle $\Theta$ between two successive bond vectors connecting its six nearest neighbors around each particle. The definition of $\Theta$, as exemplary, is illustrated in Fig.~\ref{Fig:angle_dist}-a. The definition is made such that a perfect hexagonal structure contains sharp peaks in the distribution $P(\Theta)$ at the values of $\Theta=60$, $120$, $180$, $240$, and $300$ and peaks at $0$ or $360$ are not displayed. For $\text{Pe}$ without phase separation ($\text{Pe}<55$ for $\Phi=0.2$), a flat distribution with no sharp peak indicates a homogeneous phase without long-range ordering. As anticipated, sharp peaks in the distribution of $\Theta$ spontaneously appear in Fig.~\ref{Fig:angle_dist}-a for $\text{Pe}\ge60$. Along with the increase in height, the width of the peaks also become localised to the aforementioned angles at higher $\text{Pe}$.

\begin{figure}%[h]
	\includegraphics[width=\linewidth]{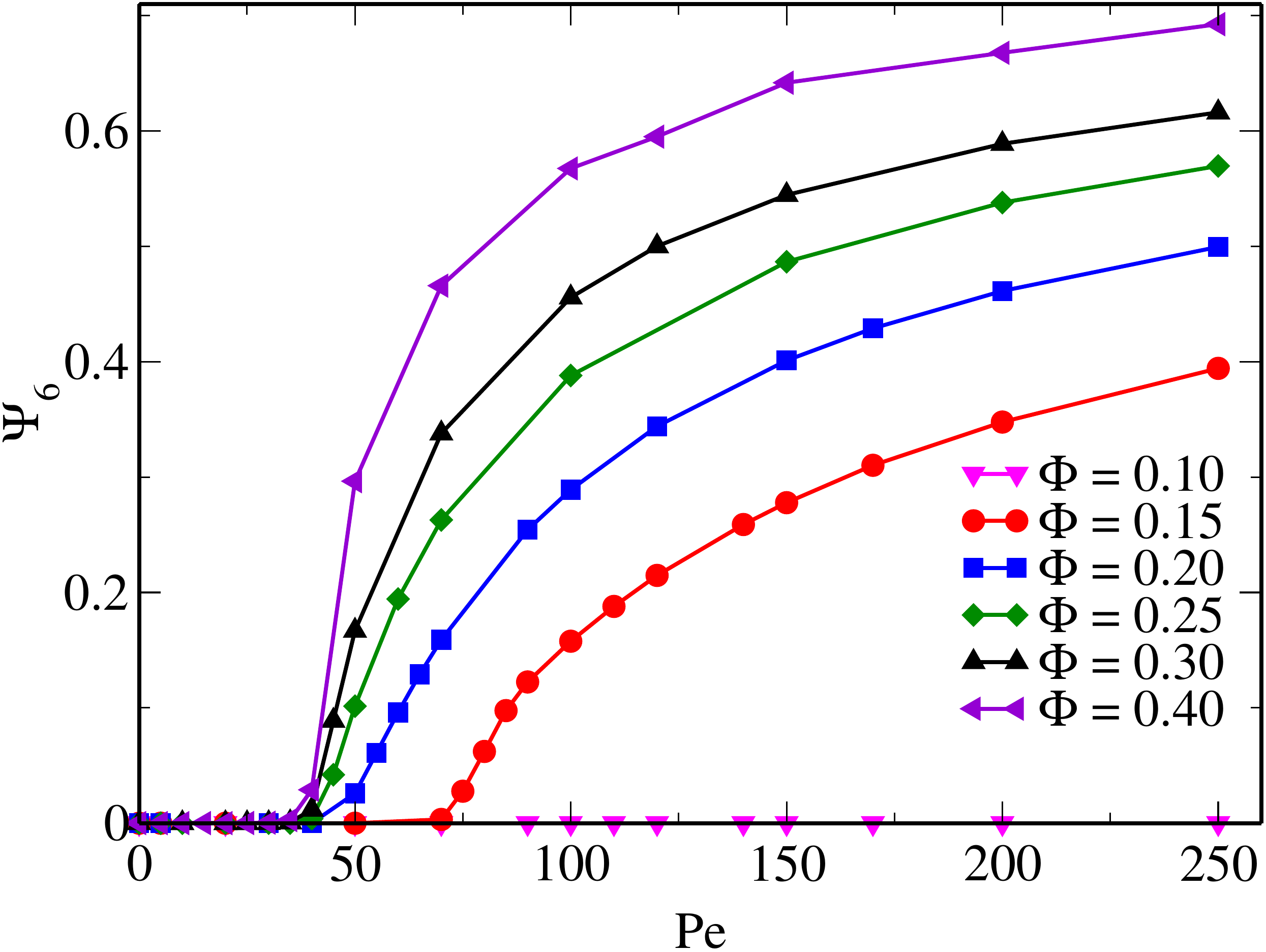}
	\caption{%a) The definition of different angles used in the analysis of the ordered phase; $\Theta$ for the angular distribution and $\theta_{ij}$ for the calculation of order parameter, $q_6$.
	The average value of $\Psi_6$  as a function P{\'e}clet number plotted for various $\Phi$. The $\Psi_6$ plot shows a steady increase with $\text{Pe}$ after the transition point indicating the growth of crystalline order within the clustered phase.}
	%b) The average orientational order parameter $\Phi$  as a function of $Pe$ for various $\phi=0.15,0.2,0.25$ and $0.3$.} 
	%The bimodal nature of the curves above $Pe\approx60$ indicates phase separation. The distribution close to $\overline{Q}_{6}=1$ having a prominent peak for $Pe>60$ confirms hexagonal close packing.}
	\label{Fig:q6_corr}
\end{figure}
A hexagonal crystalline order is prevalent in the dense clustered phase beyond $\text{Pe}_c$. To corroborate our claim of hexagonal ordering in the largest cluster in 2D, we employ a bond orientational order parameter frequently used in the literature,~\cite{theers2018clustering,bialke2012crystallization,zottl2014hydrodynamics,steinhardt1983bond,tung2016micro},
\begin{equation}
    q_{6}(i)=\frac{1}{6}\sum_{j\in n(i)}e^{i6\theta_{ij}},
\end{equation}
where $n(i)$ is the set of six nearest neighbors of $i^{\text{th}}$ particle and $\theta_{ij}$ is the bond angle between a pair $ij$ around an arbitrary axis. From the local defined parameter $q_6$, we derive a global order parameter~\cite{bialke2012crystallization},
\begin{equation}
    \Psi_6=\left<\left|\frac{1}{N}\sum_{i=1}^{N}q_{6}(i)\right|^{2}\right>.
\end{equation}
The global order parameter $\Psi_6$ lies in the range of $[0,1]$, where $0$ represents the isotropic phase of the system,  $1$ is for the perfect crystalline phase, and the intermediate value corresponds to a fraction of region with hexagonal ordering ~\cite{theers2018clustering,bialke2012crystallization,zottl2014hydrodynamics,steinhardt1983bond,tung2016micro}. The order parameter $\Psi_6$ in the considered system of quorum sensing ABPs is vanishing in the limit of small P{\'e}clet numbers ($\text{Pe}<\text{Pe}_c$). In the phase-separated state, a non-zero value of $\Psi_6$ appears with a sharp monotonic rise  as depicted  in Fig.~\ref{Fig:q6_corr}. This reveals the presence of a long-range ordered structure.   As expected, the degree of ordering increases as a function of Pe and $\Phi$.   { Analogous to $\Psi_6$, a correlation $C_{q_6}$, is generally used as a measure of local bond-order parameter to distinguish from local crystalline to  homogeneous phase. The distribution also highlights a bimodal peak in the phase-separated states (see Fig.~SI-3), which manifest local crystalline ordering.}

\begin{figure}[t]
    \includegraphics[width=\columnwidth]{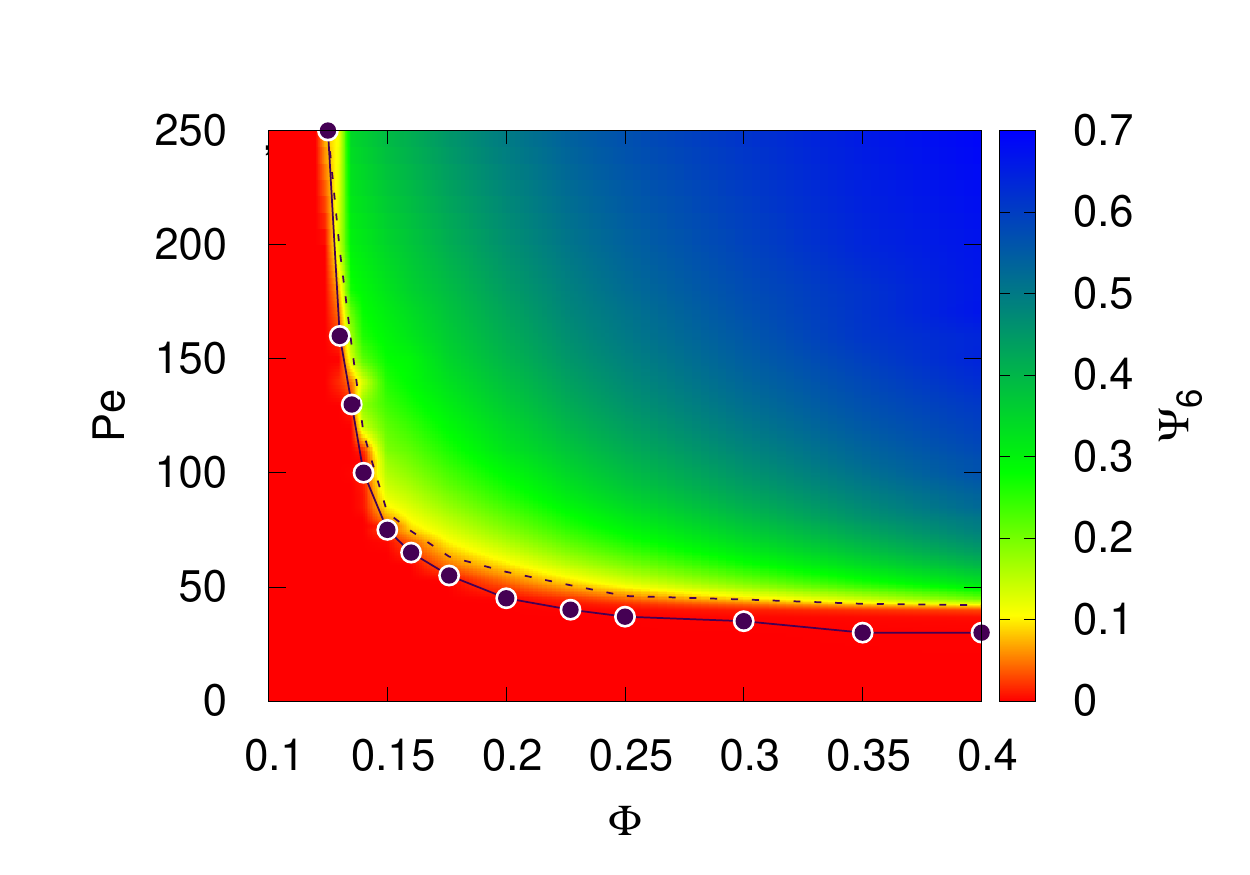}
	\caption{Phase diagram of the aggregation of active colloids in the presence of quorum-sensing in the parameter space of Pe and $\Phi$. The solid line represents  transition point ($\text{Pe}_c$) for phase separation at each $\Phi$ determined from the local density distributions. The dashed line  highlighted by a thin yellow region in the color map represents the emergence of crystalline ordering within the system ($\Psi_6\geq 0.1$) .} 
	\label{Fig:phase-diag}
\end{figure}
 The  phase-diagram of the quorum sensing ABPs are quantified and presented in Fig.~\ref{Fig:phase-diag} in the parameter space of $\text{Pe}$ and $\Phi$.   The solid line in Fig.~\ref{Fig:phase-diag} indicates the transition  P{\'ec}let number, ($\text{Pe}_c$), for which the local density distribution (Fig~\ref{Fig:clustersize}-b) reveals a bimodal curve.  The phase diagram is mapped using values of $\Psi_6$ corresponding to the color bar in Fig.~\ref{Fig:phase-diag}.
%{\cgreen A homogeneous state is represented by the predominantly red region, while the thin yellow region transitioning into green color displays the  transition boundary. The predominantly blue and green regions represent the aggregate phase.} 
A homogeneous state is represented by the predominantly red region below the solid line, while the green and blue regions represent the aggregate phase. A thin yellow region illustrates the initiation of hexagonal ordering within the dense clustered phase (also shown by a dashed line in Fig.~\ref{Fig:phase-diag}). The narrow region between the phase boundary and the dashed line marks the phase, where the aggregates are formed as a result of phase separation but  do not have a significant hexagonal ordering. The $\Psi_6$ values in this region is negligible compared to the ordered phase. The aggregate attains hexagonal ordering at slightly higher P{\'e}clet number.  At sufficiently low packing fraction, i.e., $\Phi < 0.125$, the system remains homogeneous for all Pe in the simulated window. The transition P{\'e}clet number $\text{Pe}_c$ decreases rapidly with $\Phi$ and saturates at higher values, as displayed from the solid line with bullets in Fig.~\ref{Fig:phase-diag}.

\begin{figure}[t]
\includegraphics[width=\linewidth]{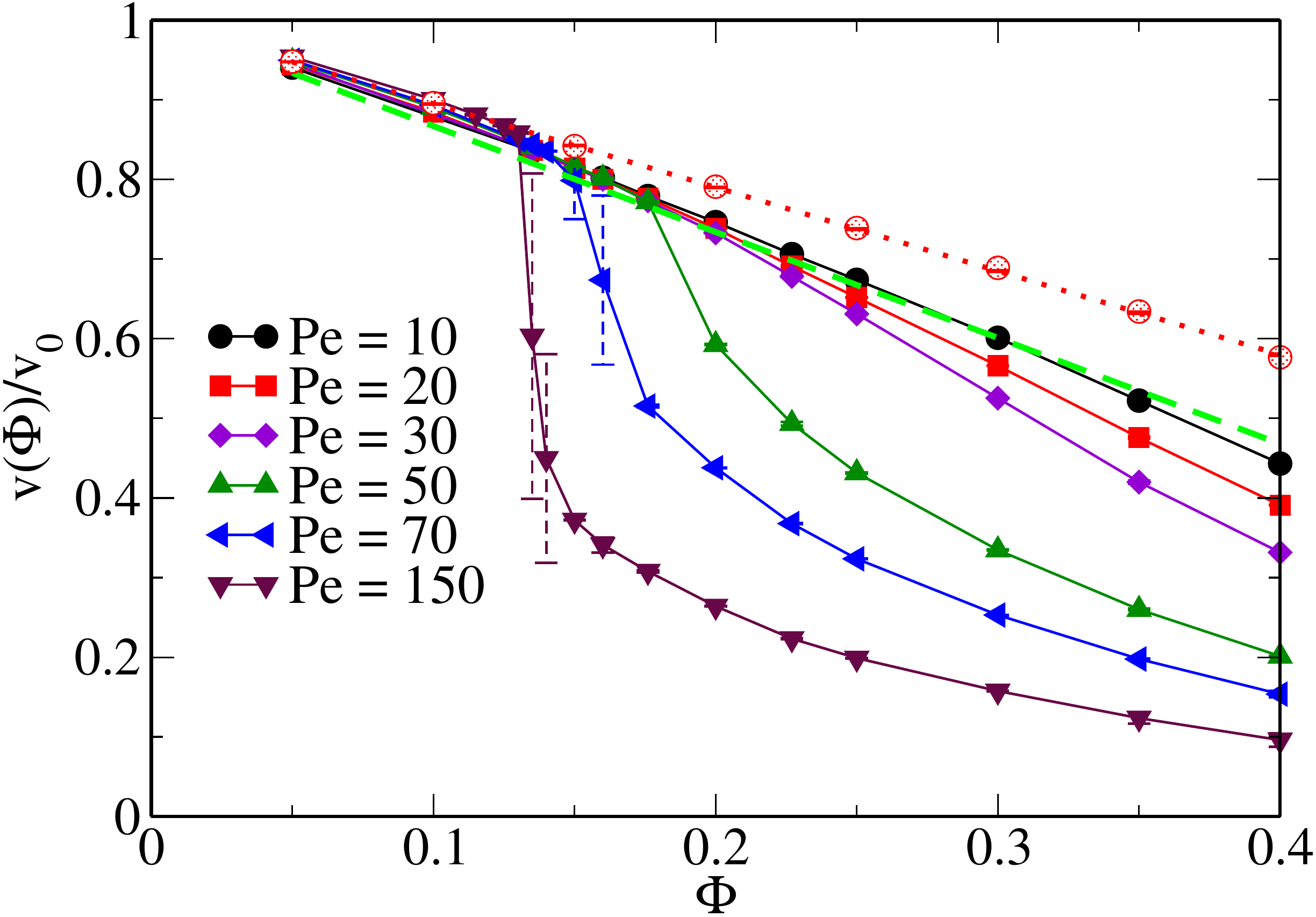}
\includegraphics[width=\linewidth]{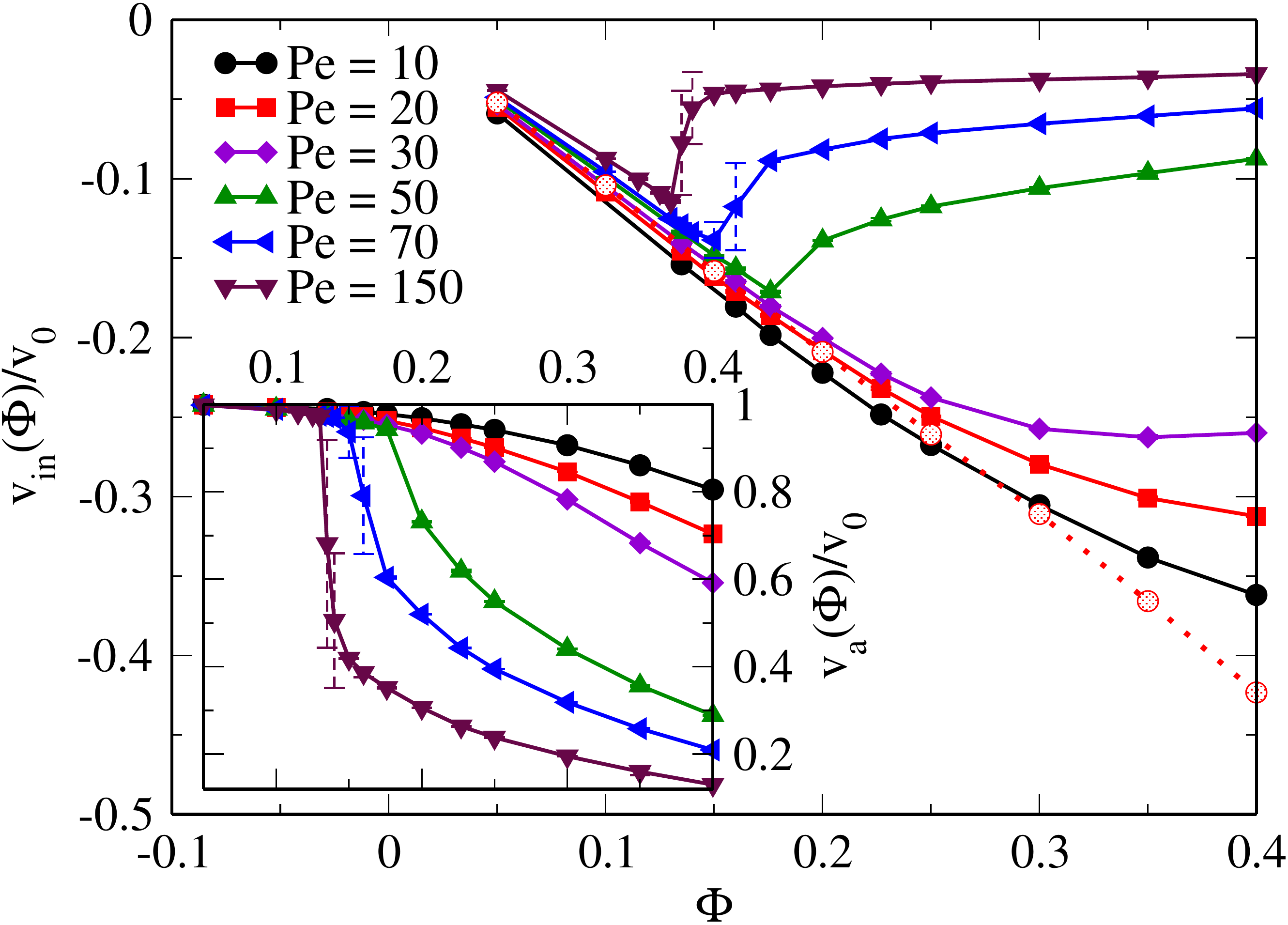}	
	\caption{
	a) The normalized swim speed as function of packing fraction $\Phi$ for various Pe. The dashed line represents a linear fit as $v(\Phi)/v_0=(1-a_q\Phi)$ with $a_q=1.34$ for $Pe=10$. The plot with red open circle  indicates the total swim speed for generic ABPs  model $\text{Pe}=30$~\cite{fily2012athermal,marchetti2013hydrodynamics}. The  red dotted line shows the linear fit as $v(\Phi)/v_0=(1-a_0\Phi)$ with $a_0=1.05$. b) Individual component of the swim speed,  interaction term $v_{in}(\Phi)/v_0$ and the active term $v_a(\Phi)/v_0$, are shown in main and inset with $\Phi$ for the same set of $\text{Pe}$. The red dotted line corresponds  for interaction contribution for generic ABPs model at $\text{Pe}=30$. The active contribution remains constant for such systems. }
	\label{Fig:local_v}
\end{figure}

 \subsection{ Swim Speed}

{ The inherent mechanism of MIPS,  without attractive or alignment interactions, involves the local slow-down of individual colloids due to an increase in local density. The local inhomogeneity near the transition point is apparent from  local density distribution in Fig.~\ref{Fig:clustersize}- a.  %Therefore, the behavior of directed swim-speed $v(\Phi)$  in the local environment  plays a vital role in the positive feedback mechanism of aggregation. % where increased collision frequency causes the its decrease, which results in further accumulation. 
The existing continuum theoretical models for MIPS is mostly based on swim speed as an input parameter; therefore, it is crucial  to obtain  this quantity from the microscopic approach ~\cite{stenhammar2013continuum,fily2012athermal}. The numerical values of  $v(\Phi)$ can be directly accessed from the simulations.  The  expression of $v(\Phi)$  is given by, 
 \begin{equation}
     v(\Phi)=\frac{1}{\gamma}\left<\Big(F_{a,i}^{'}{\bf{\hat{e}}}_i -\sum_{j=1}^{N_m}  \nabla_{i}U_{LJ}({r}_{ij})\Big)\cdot {\bf{\hat{e}}}_i\right>.
     \label{Eq:v_phi}
 \end{equation}
 The above expression can also be  written as $v=v_a+v_{in}$, where
 the first term is from the self-propulsive force $(v_a)$, and the second term $(v_{in})$ is the contribution from the interactions, which  in general is negative for repulsive potentials.  Additionally, the self-propulsion force itself is a function of local density unlike the generic ABP models where it remains a constant value~\cite{stenhammar2013continuum,fily2012athermal}. In the dilute limit $(\Phi\rightarrow0)$, the interaction contribution vanishes, and the active contribution is nearly unchanged thus $v\rightarrow v_0$, where $v_0=F_a/\gamma$ is the swim speed of an active colloids in dilute limit.

For the case of low ${\text{Pe}}$ regime, the directed speed follows approximately a linear decay with $\Phi$ (see Fig.~\ref{Fig:local_v}- a), which is in congruence with the previous observations in MIPS in ABPs~\cite{stenhammar2014phase,stenhammar2013continuum,fily2012athermal}. The linear behavior of $v(\Phi)$ weakly deviates  even in the single-phase at packing  reasonably close to the transition point even for small Pe. Notably, the deviations from linearity attain a sharp jump  with a non-linear decline in the phase-separated state as displayed in Fig.~\ref{Fig:local_v}- a for $\text{Pe}>\text{Pe}_c=45$. A noteworthy point at small $\Phi$ and Pe is that all the curves follow a universal behavior with linear variation.

It is indispensable to analyse the role of active contribution ($v_a(\Phi)$) and the interaction contribution ($v_{in}(\Phi)$) to the swim speed  separately to gain further insight. Fig.~\ref{Fig:local_v}-b shows that the magnitude of the normalised interaction term $v_{in}(\Phi)$ increases linearly with $\Phi$ for low $\text{Pe}$.   The linear behaviour of $v_{in}(\Phi)$ even in homogeneous phase is likewise of  generic ABP models at low Pe.  However, the linearity deviates at higher densities  even for low Pe in the homogeneous phase. For large $\text{Pe}>\text{Pe}_c$, $v_{in}(\Phi)/v_0$ sharply decreases with a non-monotonic  behavior. Additionally, the inset of Fig.~\ref{Fig:local_v}- b displays the normalised active contribution, $v_a(\Phi)/v_0$. Here, all curves follow a universal decreases  at low densities. The deviation from this universal behavior appears at small Pe,  for high densities, even for $\text{Pe}=10$. The deviation displayed for $\text{Pe}\geq 20$ is congruent with the behavior of $v(\Phi)$ in Fig.~\ref{Fig:local_v}- a.  Moreover, near the  phase-separation ($\text{Pe}>\text{Pe}_c$), $v_a(\Phi)$ also consist a sharp  drop similar to $v(\Phi)$  Fig.~\ref{Fig:local_v}-a.
 It reveals that  a drastic  change in $v_a$ is predominately responsible for the sharp decrease in the directed speed  $v(\Phi)$ at larger $\text{Pe}$. Here, the range of quorum-sensing  $R_{c}=1.3\sigma$ is larger than the  short-range LJ interaction. This causes the speed to sharply decrease even before they undergo direct pair-wise collision. Thereby, as expected the slow moving particles will have smaller interaction values of $v_{in}$ to that of fast moving particles. Accordingly, the impact of collision contribution becomes weaker, hence  $v_{in}$ for denser system sharply diminishes which is captured in Fig.~\ref{Fig:local_v}-b main plot,  at $\text{Pe}=50,70,$ and $150$ ).

Furthermore, we have computed local swim speed with local packing fraction $\Phi_l$, which is illustrated in Fig. SI-6. The behavior is very similar to the globally averaged quantity, which additionally corroborates the claim of local slow-down at the microscopic level.  To summarise,   multi-body collisions   emerges at higher $\text{Pe}$ and $\Phi$, causing to set in the quorum-sensing behavior, which  plays dominant role  in  the reduction of the directed speed unlike  the generic ABP model where interaction term is vital. }

\subsection{Dynamics of cluster}
\begin{figure}[t]
	\includegraphics[width=\linewidth]{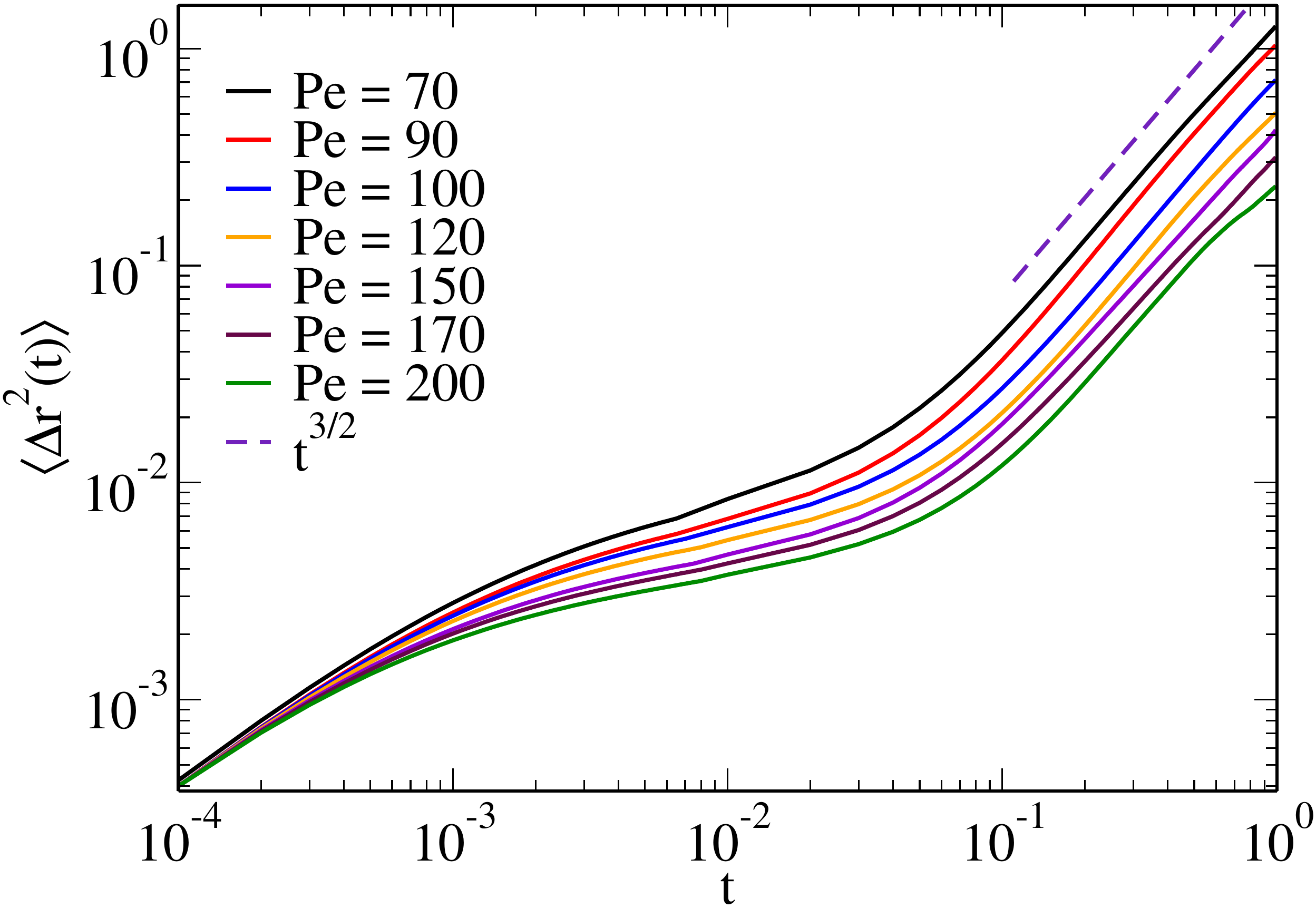}
	\includegraphics[width=\linewidth]{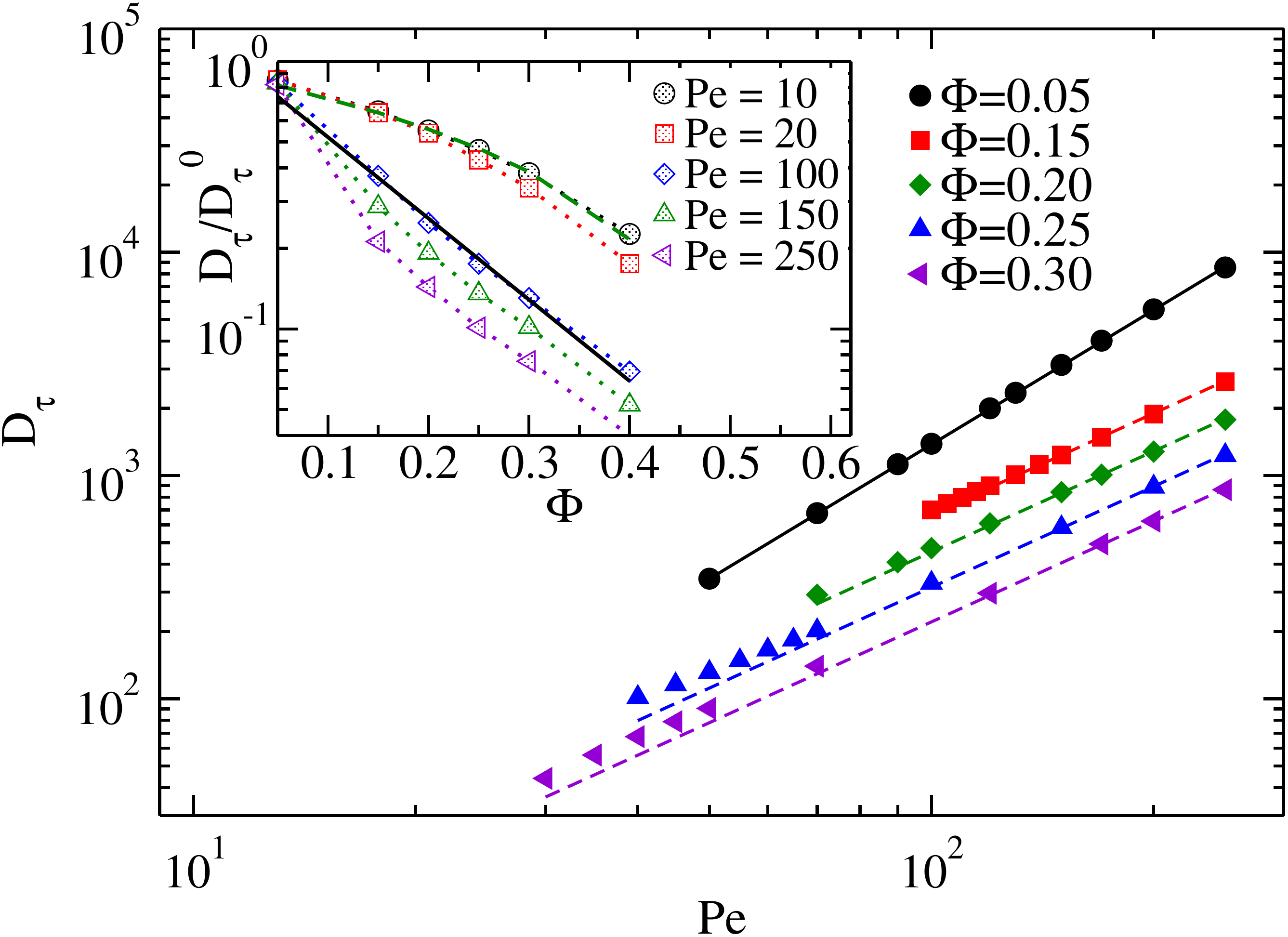}
	\caption{a) The  MSD of particles deep inside the cluster  for different $\text{Pe}$. The dashed line shows the super-diffusive behavior as $t^{3/2}$.
	b) The effective diffusion coefficient as a function of P{\'e}clet number for $\Phi=0.05$ to $\Phi=0.4$. The solid line shows power-law behavior, $D_\tau\approx\text{Pe}^2$, and dashed lines as $D_\tau \approx\text{Pe}^{3/2}$. The inset shows normalized $D_\tau/D_{\tau}^{0}$ as a function of $\Phi$ for various $\text{Pe}$.  The solid line displays exponential fit as $D_\tau/D_\tau^{0}\sim e^{- b_2\Phi}$, with $b_2\approx 7.0$, while the  dashed line displays linear fit as $D_\tau/D_\tau^{0}\sim 1-b_1\Phi$ with $b_1=1.9$.}
	\label{Fig:msd}
\end{figure}

The dynamical behavior of a colloid in the aggregate can be assessed through its mean-square-displacement (MSD). At the boundary of the aggregate, the colloid moves  in and out from the cluster. Therefore,  a convenient  way to compute the MSD of the clustered particles is to consider those particles which are close to the centre of the cluster so that they do not leave the cluster in the characterised time window.
As expected, at short time scales, the particles exhibit sub-diffusive behavior due to the constraint from the presence of several neighboring colloids in its proximity (see Fig.~\ref{Fig:msd}-a). At intermediate time scale, the self-propulsion is   suppressed as the particles are caged within the aggregate. Consequently, MSD attains a plateau at this timescale. More importantly, the width of the plateau widens and becomes more prominent at higher $\text{Pe}$,  as Fig.~\ref{Fig:msd}-a illustrates for  a range of $\text{Pe}=70$ to $200$. In the long time limit, colloid attains diffusive behavior.  The enhanced diffusive dynamics of colloids are evident from MSD curves in  the timescale $t>0.1\tau$. A dashed line indicates the super-diffusive behavior, where the MSD is super-diffusive according to $<\Delta r^{2}(t)>\sim t^{3/2}$ with an exponent $3/2$. The super-diffusive behavior is a consequence of large active force on the surface of the cluster, which drives significant fluctuations in its shape and additionally the  propagation of fast-moving  defects inside the cluster. 
{ The defect dynamics  and shape of the cluster  is illustrated  in SI-Movie-1 and 2 for visualisation.} 
%{\cred Thus, the microscopic effects which transpire inside the cluster driven by the surface effects conclusively govern  overall dynamics of the clustered phase.} 

The characteristic dynamical  features of the colloids can be presented  in terms of the effective diffusion coefficient. For this, the MSD of the colloids is averaged over all particles (Fig. SI-4). % The MSD of the active colloids exhibits  super-diffusive (intermediate-time) followed by the diffusive  regime (long-time). 
  The measured effective diffusion coefficient, from the diffusive regime, is plotted as a function of $\text{Pe}$ in Fig.~\ref{Fig:msd}-b. For $\Phi=0.05$,  the effective diffusion coefficient increases quadratically with $\text{Pe}$. On contrary, for  $\Phi \ge 0.125$ and $\text{Pe} > \text Pe_c$,  $D_\tau$  grows with a power of exponent $3/2$ with  Pe (see Fig.~\ref{Fig:msd}-b). The effective diffusion coefficient shows a cross-over from $\text{Pe}^{2}$ quadratic growth  in the homogeneous state to $\text{Pe}^{3/2}$ behavior in the phase-separated state. 

The inset of Fig.\ref{Fig:msd}-b shows behavior of the normalised $D_{\tau}/D_{\tau}^{0}$, where $D_{\tau}^{0}=D_t(1+\frac{1}{6}\text{Pe}^2)$ is the effective diffusion coefficient for ideal active particles at the same $\text{Pe}$~\cite{winkler2015virial,swim_kardar}. As expected from the effective diffusion coefficient, the diffusive dynamics of particle substantially slows down with $\Phi$, as $D_\tau$ decreases with $\Phi$ for all $\text{Pe}$~\cite{winkler2015virial}. { In the homogeneous phase, the effective diffusion coefficient follows approximately a linear behaviour, $D_\tau(\Phi)/D_\tau^{0}\sim 1-b_1\Phi$ with $b_1\approx1.9$ with a larger slope reported in  previous studies~\cite{stenhammar2014phase,stenhammar2013continuum,fily2012athermal}. However, for $\text{Pe}>\text{Pe}_c$, the diffusion is suppressed  sharply, as result shows an exponential behavior as,  $D_\tau(\Phi)/D_\tau^{0}\sim\exp(-b_2\Phi)$ with $b_2\approx7.0$. }%The exponential  decay of $D_\tau$, with $\Phi$ for large $\text{Pe}$,  gives a justification (along with the slowdown in swim speed discussed before) for the phase separation for the density $\Phi\ge0.125$ at large $\text{Pe}$. Moreover, the scaled effective diffusion further decreases with increase in $\text{Pe}$. } 
\begin{figure}[t]
	\includegraphics[width=\linewidth]{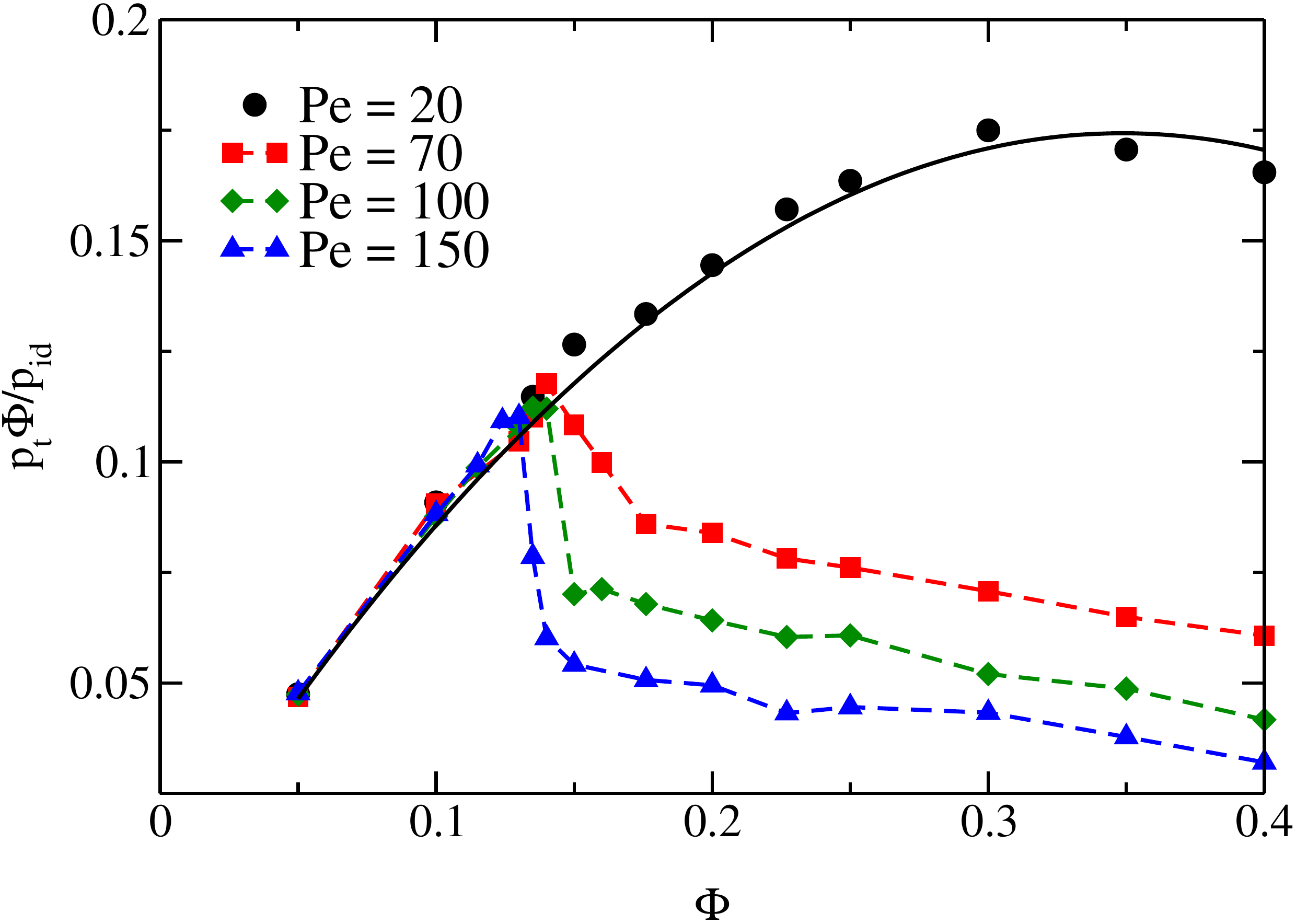}
	\caption{Total pressure $p_t$ plotted as a function of $\Phi$ for various $\text{Pe}=20$, $70$, $100$ and $150$. Solid line shows the function, $p_t= \phi(1-\kappa \Phi)$, where $\kappa\approx1.42$.}% The behavior of n swim pressure with solid line $Pe^2$ and dashed line $Pe^{3/2}$.  } 
	\label{Fig:swim_press}
\end{figure}

\subsection{Pressure}
The total pressure of a system undergoing MIPS is used to identify phase boundaries as it exhibits a non-monotonic behavior on the density with a sharp jump at the transition point~\cite{winkler2015virial,digregorio2018full}. The total pressure exhibits contributions from the thermal fluctuations, the interaction force, and the active force (active pressure). Total pressure can be expressed as $p_t=p_0+p_a+p_{in}$, with individual contributions as, $p_0=k_BT/A$, the ideal gas pressure in 2D, $p_a$, the active pressure, and $p_{in}$, the standard virial term due to inter-particle interactions. The explicit forms of $p_a$ and $p_{in}$ are given as,

\begin{eqnarray}\nonumber
    p_a&=&\frac{\gamma}{2AN}\sum_i^N\left<F_{a,i}^{'}\mathbf{\hat{e}}_i.\mathbf{r}_i\right>, \\ 
    p_{in}&=&\frac{1}{4AN}\sum_{j,i}^{N}\left<\mathbf{F}_{ij}\cdot(\mathbf{r}_i-\mathbf{r}_j)\right>.	
\end{eqnarray}
 The total pressure $p_t(\Phi)$ defines an equation of state of ABPs without confinement. The ideal swim pressure is given as $p_{id}A=k_BT(1+\frac{1}{6}\text{Pe}^{2})$ at a temperature $T$ in two dimensions~\cite{winkler2015virial,swim_kardar}. This expression  acts as pressure for the ideal active particles at any density. %This relation of interacting colloid is also valid for low-density $\Phi$.

 %In the limit of large $\text{Pe}$, the power-law exponent for the active pressure is the same as for the effective diffusion coefficient as suggested in Ref.~\cite{winkler2015virial,Brady_swim_prl2014}. 
 
 %{\cred The active term $p_a$ always dominates in the total pressure over the interaction contribution. In the dilute limit,  the inter-molecular interactions diminish with a very nominal contribution. However, in the phase separated state, active force dominates on the surface of the clustered phase (see Fig. SI-7). The contribution of the surface effects is thus very prominent in the global dynamics of the clustered phase and is reflected in the active pressure.}

Figure~\ref{Fig:swim_press} presents the   scaled total pressure with $p_{id}$ multiplied by $\Phi$.  Importantly, $p_t \Phi/p_{id}$ shows a universal  behavior until the onset of phase separation. In the single-phase, we can write  pressure as $p_{t}\Phi/p_{id}=\Phi(1-\kappa\Phi)$ with $\kappa =1.42$. A sudden drop in pressure occurs at the transition point~\cite{winkler2015virial,Brady_swim_prl2014,solon2015phase} in Fig.~\ref{Fig:swim_press}. It suggests a discontinuous drop in active particles' pressure due to slowdown of their speed in clusters.  This behavior is in contrast to the plot of a continuous parabolic profile for $\text{Pe}=20$, where  system exist in  a homogeneous single phase. The change in the  behavior of pressure in the phase-separated state is a consequence of large-number fluctuations in the active systems~\cite{fily2012athermal} (see Fig.~SI-1). 
It is important to emphasize here that relatively more pronounced drop in pressure is noticeable  at the transition point than the previous studies~\cite{winkler2015virial,Brady_swim_prl2014,solon2015phase}.  
In phase separated state, active contribution dominates   (see Fig. SI-5), which shows that the surface effects are more prominent in the behavior of global variables.

%The active pressure reflects the surface contributions and super-diffusivity of the clustered particles, and total pressure (Fig.~\ref{Fig:swim_press}) describes the phase-boundaries, the description of pressure effectively summarises the overall global dynamics of our system (see Fig. SI-7).

%The change in the behavior of the pressure in the phase separated state is a consequence of large-number fluctuations in the active systems.

\subsection{Non-vanishing Motility of Quorum-Sensing}
\begin{figure}%[h]
    \hspace*{-0.5cm}
	\includegraphics[width=\linewidth]{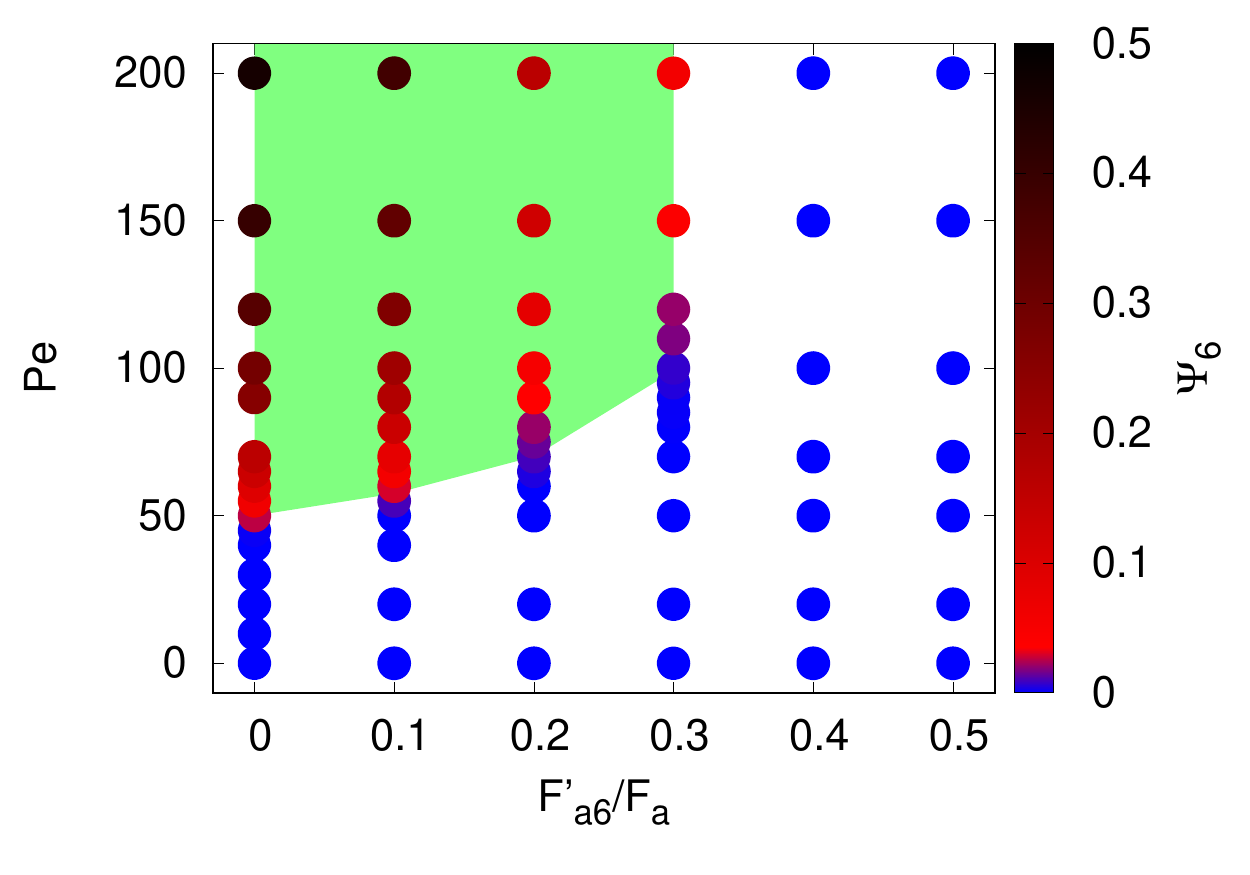}
	\caption{Phase diagram of the aggregation of active colloids in the presence of quorum-sensing in the parameter space of Pe and $F_{a6}^{'}/F_a$ at $\Phi=0.2$.} 
	\label{Fig:non-zero speed}
\end{figure}

 All  the results so far discussed corresponds to the particular case  where the active force becomes zero whenever a colloid gets surrounded by six or more colloids as defined in Eq.~\ref{Eq:qur}.  Now we test our approach to a broader class of problem by mimicking non-zero motility in the dense region for $n\ge 6$.  The mathematical equation for the density-dependent motility  is same as given in Eq.~\ref{Eq:qur}, while the choice of $\alpha$ and $\beta$ dictates a non-zero active force at $n \ge 6$, and are listed in table I. A series of simulations are carried out for a range of $F_{a6}^{'}/F_{a}$, where $F_{a6}$ is the active force at $n\ge6$, to test its influence on the phase-separation and robustness of the choice of the approach.  The summary of  results   are translated into a phase diagram in Fig.~\ref{Fig:non-zero speed}, where it displays the aggregation of colloids in the parameter space of $\text{Pe}$ and ratio of $F_{a6}^{'}/F_{a}$ at $\Phi=0.2$. The shaded area in the plot  displays phase separated state with hexagonal ordering. It  indicates that the system continues to retain phase separation up to $F_{a6}^{'}/F_a \le 0.3$. As expected, phase-separation disappears for larger residual motility, and system attains the homogeneous phase further for $\Phi=0.2$.

\begin{table}%[h!]
	\begin{center}
		\caption{Values of $\alpha$, $\beta$, and  strength of active force  $F_{a6}^{'}/F_a$  at $\Phi=0.2$.  Here $F_{a}^{'}=F_{a6}^{'}$ corresponds to the values of active force  for $n \geq 6$. }
		\label{tab:table1}
		\begin{tabular}{|c|c|c|}
		\hline
			\hspace{0.5cm} $\alpha$ \hspace{0.5cm} & \hspace{0.5cm} $\beta$ \hspace{0.5cm} & \hspace{0.5cm} $F_{a}^{'}/F_{a}$ \hspace{0.5cm}  \\
			\hline
			\hspace{0.5cm} 1.0 \hspace{0.5cm} & \hspace{0.5cm} 1.8 \hspace{0.5cm} & \hspace{0.5cm} 0.0 \hspace{0.5cm} \\
			\hspace{0.5cm} 1.06 \hspace{0.5cm} & \hspace{0.5cm} 1.62 \hspace{0.5cm} & \hspace{0.5cm} 0.1 \hspace{0.5cm} \\
			\hspace{0.5cm} 1.14 \hspace{0.5cm} & \hspace{0.5cm} 1.44 \hspace{0.5cm} & \hspace{0.5cm} 0.2 \hspace{0.5cm} \\
			\hspace{0.5cm} 1.24 \hspace{0.5cm} & \hspace{0.5cm} 1.26 \hspace{0.5cm} & \hspace{0.5cm} 0.3 \hspace{0.5cm} \\
			\hspace{0.5cm} 1.37 \hspace{0.5cm} & \hspace{0.5cm} 1.08 \hspace{0.5cm} & \hspace{0.5cm} 0.4 \hspace{0.5cm}  \\
			\hspace{0.5cm} 1.55 \hspace{0.5cm} & \hspace{0.5cm} 0.8 \hspace{0.5cm} & \hspace{0.5cm} 0.5 \hspace{0.5cm}  \\
				\hline
		\end{tabular}
	\end{center}
\end{table}

\section{Summary}
We have presented over-damped Langevin dynamics simulations of active colloids whose motility vary with their local density.  The alteration of self-propulsion speed with density is observed in nature  for example  bacterial systems~\cite{sokolov2018prl}.  
A local model of quorum-sensing has been adopted in the past for the study of collective phenomena in active systems, and it displays fascinating  global  dynamics~\cite{velasco2018collective,mishra2012collective,farrell2012pattern,fischer2020quorum}. 
Our study reports that the system phase separates into a dense phase coexisting with a low-density phase at very small packing fractions ($\Phi\ge 0.125$)  ~\cite{reentrant_baskaran,bialke2015active,bialke2012crystallization}. The dense clustered phase acquires hexatic ordering at higher activity strength beyond the critical point. { Active particles in the cluster exhibit sub-diffusive dynamics at a short time due to local caging from neighboring particles. 
%Further,  quorum-sensing colloids display a similar relation between active pressure and effective diffusion coefficient.  
The effective diffusion coefficient grows according to  $\text{ Pe}^{3/2}$  in the phase-separated state,~\cite{winkler2015virial,Brady_swim_prl2014} while a parabolic increase in the homogeneous phase is reported.}

It has previously been established that MIPS is a consequence of a positive feedback mechanism between accumulation induced slow down, and slow speed drives accumulation in ABP models~\cite{Cates_2013,tailleur2008statistical}. In our study, the mechanism behind aggregation is instigated by quorum-sensing phenomenon due to emergence of multi-body collisions at higher density and Pe.  This sets up an apparent slow-down causes a sharp decline in the relative directed speed.
Further, we have revealed  that the slow down is dominated by quorum-sensing contribution (see Fig.\ref{Fig:local_v}-b). Notably,  the role of interaction contribution is weaker in contrast to the generic models.   Slow down of speed also accounts for the reduction in effective-diffusivity, facilitating a crossover from the linear to exponential behavior. 
 
 The phase-separation strongly depends on the choice of the $R_c$, a larger  $R_c$ will push the phase-separation at slower  self-propulsion speeds or smaller packing fractions.  A smaller values will lead our model towards the standard ABP model, and the phase-separation at lower packing disappears.  Additionally, it can also be visualized in  simulations that the density-dependent  contribution is sufficient enough to bring the phase-separation, likewise in the theoretical studies~\cite{solon2015phase}, at relatively small P{\'e}clet numbers. Besides, these  phase-separated states lag the long-range ordering  present in largest cluster  without pair-wise repulsive force (see Figure SI-7). 
 
 The  linear behavior of $v(\Phi)$ and $D(\Phi)$ in single phase with existing models~\cite{fily2012athermal,solon2015phase} exhibits the coefficients $a_q=1.34$ and $b_1=1.9$ larger  than the  previously reported  approach~\cite{Cates_2013,tailleur2008statistical,fily2012athermal}.  The  Eq~.\ref{Eq:qur} suggests   quorum-sensing influence causes  a bigger effective diameter  of colloidal interaction  ($R_c=1.3$)  than  the  generic repulsive ABP models, consequently they exhibit a larger  effective scattering cross-section. This attributes to  steeper slope ($a_q$ and $b_1$) in the reduction of directed speed and effective diffusion coefficient.

%This early deviation is a primary indicator of how the microscopic dependence of quorum sensing perturbation enhances the rate of aggregation by arresting motility of caged colloids.
%Quorum sensing introduces a microscopic dependence of the the local density on the self-propulsion force. The addition of the quorum sensing perturbation makes the effective scattering cross-section of colloidal aggregates bigger than that for generic repulsive ABP model. x

%Hereby, assuming the effective diameter $R_c=1.3$, it will cause increase of the effective cross-section by a factor of $1.69$.

%This is by  virtue of a complex relationship between mean collision time, on P{\'e}clet number and local density. The dependence of $a_q$ on $Pe$ is more apparent at larger  P{\'e}clet numbers where velocity is non-linear function of local $\Phi_l$ (see Figure.\ref{Fig:local_v}).

%The best fit for the velocity in the single phase at $Pe=30$ leads to $a_q=1.38$.  

% In a system where local effects are relevant, it is imperative that the microscopic dynamics has to be addressed in parallel to complete the picture. 
The presented results are not unique for the adopted linear model proposed in Eq.~\ref{Eq:qur}; an exponentially decaying self-propulsion force also yields a similar phase-separation nearly at the same packing fraction. The qualitative behavior is presented in the supplementary text; see Fig.~SI-7, and 8.
In conclusion, a  microscopic model of local density-dependent motility plays a crucial role in the kinetics of MIPS at low density, and the same mechanism also facilitates phase separation.  The reduction in local directed speed (see SI-Fig. 6) of colloids asserts the importance of probing  the microscopic dynamics, it may be useful in providing a complete picture of phase-diagram. Our  approach considers  simpler interactions; however  various other effects may play a vital role in  the collective dynamics. For example, viscous lubrication, hydrodynamics, and anisotropic shape of colloids may be worth considering for the realistic models. Additionally, a model  of coarse-grained  density  (where local density is computed in weighted averaged fashion)  over microscopic model might serve a better purpose in direct comparison with theoretical results~\cite{solon2018generalized}.

\section*{Acknowledgements}
The authors would like to thank the HPC facility at IISER Bhopal for the computation time and  DST SERB Grant No. YSS/2015/000230 and CRG/2020/000661 for the financial support. SKA and FJ  thank Harsh Kumar for indulging in useful discussions at the initial stage of the work.

%%%END OF MAIN TEXT%%%

%%%REFERENCES%%%
%\bibliography{rsc} %You need to replace "rsc" on this line with the name of your .bib file
\bibliographystyle{rsc} %the RSC's .bst file

\providecommand*{\mcitethebibliography}{\thebibliography}
\csname @ifundefined\endcsname{endmcitethebibliography}
{\let\endmcitethebibliography\endthebibliography}{}

\end{document}